\documentclass[superscriptaddress,amsmath, nofootinbib]{revtex4}
\usepackage{amsfonts}
\usepackage{graphicx}
\usepackage[brazil]{babel}
\usepackage[latin1]{inputenc}
\usepackage{amsmath}
\usepackage{amssymb}
\begin{document}
%\draft

%\preprint{APS/123-QED}

\title{Dynamics on a submanifold: intermediate formalism versus Hamiltonian reduction of Dirac bracket, and integrability.}
%{Intermediate Hamiltonian formalism and Poisson 
%geometry of a particle confined to move on a submanifold.}

\author{Alexei A. Deriglazov }
\email{alexei.deriglazov@ufjf.br} \affiliation{Depto. de Matem\'atica, ICE, Universidade Federal de Juiz de Fora,
MG, Brazil} 
%\affiliation{Department of Physics, Tomsk State University, 
%Lenin Prospekt 36, 634050, Tomsk, Russia}

\author{}%{Guilherme Corr\^ea Silva }
%\email{guilherme.jfa@hotmail.com} \affiliation{Depto. de F\'isica, ICE, Universidade Federal de Juiz de Fora, MG,
%Brazil}

\date{\today}% It is always \today, today,

\begin{abstract}
We consider Hamiltonian formulation of a dynamical system forced to move on a submanifold $G_\alpha(q^A)=0$.  If for some reasons we are interested in knowing the dynamics of all original variables $q^A(t)$, the most economical would be a Hamiltonian formulation on the intermediate phase-space submanifold spanned by reducible variables $q^A$ and an irreducible set of momenta $p_i$, $[i]=[A]-[\alpha]$. We describe and compare two different possibilities for establishing the Poisson structure and Hamiltonian dynamics on an intermediate submanifold: Hamiltonian reduction of the Dirac bracket and intermediate formalism. As an example of the application of intermediate formalism, we deduce on this basis the Euler-Poisson equations of a spinning body, establish the underlying Poisson structure, and write their general solution in terms of the exponential of  the Hamiltonian vector field.  
\end{abstract}

\maketitle %\noindent
%%%%%%{\bf DOI:}
%%%%%%%{\bf PACS numbers:} 11.10.Ef, 03.65.Ca \\
%%%%%%%%{\bf Keywords: Thomas Precession, Relativistic Spin, Noncommutative Geometry}

%\tableofcontents

%\newpage

\section{Three equivalent Hamiltonian formulations for a system with holonomic constraints.}\label{CAP1.1}

Consider a mechanical system that can be described with help of non-singular Lagrangian $L(q^A, \dot q^A)$, defined on configuration space with generalised coordinates $q^A(t)$, $A=1, 2, \ldots n$.  Suppose the "particle" ~$q^A$ was then forced to move on a $k$\,-dimensional surface ${\mathbb S}^k$ given by the algebraic  equations $G_\alpha(q^A)=0$. The task is to construct the Hamiltonian formulation for this theory.  There are three different possibilities to do this. Let us first shortly describe and compare them.

{\bf (A)} The first possibility is to work with unconstrained variables. Let $x^i$, $i=1, 2, \ldots , k$ be local coordinates on ${\mathbb S}$.   Then equations of motion follow from the Lagrangian $\tilde L(x^i, \dot x^i)\equiv L(q^A(x^i), dq^A(x^i)/dt)$. If $\tilde L$ is also non-singular, we introduce the conjugate momenta  $p_i$ for $x^i$, the Hamiltonian $H(x^i, p_j)$, and the canonical Poisson 
bracket $\{x^i, p_j\}=\delta^i_j$.  Then the Hamiltonian equations are $\dot x^i=\{x^i, H\}, \dot p_i=\{p_i, H\}$. 

The transition to independent variables $x^i$ is not always desirable. For instance, in the case of a spinning body, the $q^A$ variables are 9 elements of orthogonal $3\times 3$ matrix $R_{ij}$ (therefore $G_\alpha=0$ reads as $R^TR-{\bf 1}=0$). To describe a rigid body, we need to know the evolution of $q^A$ and not $x^i$.

{\bf (B)} The second possibility is to work with original variables using the Dirac's version of Hamiltonian formalism \cite{Dir_1950, GT, deriglazov2010classical}.  Equations of motion follow from the modified Lagrangian action, where the constraints are taken into account with help of auxiliary variables $\lambda_\alpha(t)$  as follows \cite{Arn_1, deriglazov2010classical}:  
\begin{eqnarray}\label{int02}
S=\int dt ~ L(q^A, \dot q^A)-\lambda_\alpha G_\alpha(q^A). 
\end{eqnarray}
We should pass to the Hamiltonian formulation introducing the conjugate momenta $p_A, p_{\lambda\alpha}$ to all original variables $q^A, \lambda_\alpha$. The Hamiltonian equations then obtained with help of canonical Poisson brackets $\{q^A, p_B\}=\delta^A{}_B$, $\{ \lambda_\alpha,  p_{\lambda \beta}\}=\delta_{\alpha\beta}$, and with help of Hamiltonian of the form $H(q^A, p_B, \lambda_\alpha, p_{\lambda \beta})$. The resulting equations depend on the auxiliary variables $\lambda_\alpha$ and $p_{\lambda\alpha}$.   The systematic method for excluding them is to pass from the canonical to Dirac bracket. The latter is constructed with help of second-class constraints 
\begin{eqnarray}\label{int02.02}
G_\alpha(q^A)=0, \qquad \Phi_\alpha(q^A, p_B)=0,
\end{eqnarray}
that appear in the Hamiltonian formulation of the theory (\ref{int02}). Working with the Dirac bracket, all terms with auxiliary variables disappear from the final equations. This gives Hamiltonian formulation on the phase space with coordinates $q^A, p_B$.  

{\bf (C)}  In the case of a spinning body, a kind of intermediate formulation arises between (A) and (B). The freely spinning body can be described by  $9+3$ Euler-Poisson equations\footnote{By $({\bf a}, {\bf b})$ and $[{\bf a}, {\bf b}]$ we denote the scalar and vector products of the 
vectors ${\bf a}$ and ${\bf b}$.}
\begin{eqnarray}\label{int01.1} 
\dot R_{ij}=-\epsilon_{jkm}\Omega_k R_{im}, , \qquad 
I\dot{\boldsymbol\Omega}=[I{\boldsymbol\Omega}, {\boldsymbol\Omega}], 
\end{eqnarray}
where $I$ is a numerical $3\times 3$ matrix. They turn out to be the Hamiltonian equations \cite{Chet_1941,Mar_98,Hol_2007,AAD23_1,AAD23}, with the configuration-space variables assembled into a $3\times 3$ matrix $R_{ij}(t)$, while $\Omega_i(t)$ are three components of momenta. There are 9  redundant coordinates $R_{ij}$, but only 3 independent momenta $\Omega_i$. So, if in case (A) we worked with unconstrained set $(x^i, p_j)$, and in case (B) with redundant set $(q^A, p_B)$, then now we have an intermediate situation: $(q^A, p_j)$. This gives the most economical Hamiltonian formulation of a theory in which we are interested in knowing the dynamics of all variables $q^A$. 

An intermediate formulation for the theory (\ref{int02}) can be obtained in the Dirac's formalism, by first constructing the Dirac bracket (which is a degenerate Poisson structure on original phase space $(q^A, p_B)$), and then reducing it on the submanifold $\Phi_\alpha=0$. Let us call it the intermediate submanifold\footnote{All solutions of the theory (\ref{int02}) lie in the phase-space submanifold $\Phi_\alpha=0$, $G_\alpha=0$, hence the term "intermediate".}. In the present work we develop an alternative way, allowing to construct the Poisson structure on this submanifold without the need for the Dirac bracket. 
Roughly speaking, this works as follows. For any theory of the form (\ref{int02}) with positive-definite Lagrangian $L$, we present an universal procedure to find (non-canonical) phase-space coordinates $(q^A, \pi_i, \pi_\alpha)$ with special properties. They are constructed with help of the matrix $G_{\alpha A}\equiv\partial G_\alpha/\partial q^A$ and fundamental solutions of the linear system $G_{\alpha A}x_A=0$. The intermediate formulation of the theory (\ref{int02}) is obtained by first rewriting the Hamiltonian formulation of unconstrained theory $L$ in terms of new coordinates, and  then excluding  the variables $\pi_\alpha$ from all resulting expressions with help of the constraint $\Phi_\alpha=0$. In particular, the Poisson structure on intermediate submanifold turns out to be the canonical Poisson bracket of original variables $(q^A, p_B)$, first rewritten in terms of new coordinates $(q^A, \pi_B)$, and then restricted to this submanifold.

As we saw above, an interesting application of the intermediate formalism lies in the branch of a spinning body dynamics. These issue is also of interest in modern studies of various aspects related with construction and behaviour of spinning particles and rotating bodies in external fields beyond the pole-dipole approximation \cite{Dan_2023, Bena_2023, Sour_2023, Armen_2022, Hwi_2023, Yass_2023, Beck_2023, Farr_2023, Fau_2023}. For simple mechanical systems (point particle in an external field or several mutually interacting particles), their equations of motion are postulated on the base of analysis of experimental data. Unfortunately, a spinning body turns out to be too complex system to find its equations in this way. So,  even writing the equations of motion of a spinning body turns out to be a non-trivial task. At the dawn of the development mechanics, this was considered as one of the central problems, for which several branches of classical mechanics were developed: Lagrangian mechanics on a submanifold, Hamiltonian mechanics with constraints, symmetry groups and their relation with conservation laws and integrals of motion, integrable systems and so on.  
In the result, the basic theory of a rotating body was formulated in the works of Euler, Lagrange, Poisson, Poinsot and many others \cite{Eul_1758, Lag_1788, Poi_1842, Poin}. However, a didactically systematic formulation and application of these methods to various problems of  rigid body dynamics is still regarded not an easy task \cite{Mar_98, Hol_2007}. For instance, J. E. Marsden, D. D. Holm and T. S. Ratiu in their work \cite{Mar_98} dated by 1998 write: "It was already clear in the last century that certain mechanical systems resist the usual canonical formalism, either Hamiltonian or Lagrangian, outlined in the first paragraph. The rigid body provides an elementary example of this."

Second-order Lagrangian equations of a spinning body can be obtained as the conditions of extreme of a variational problem, where the body is considered  as a system of particles subjected to holonomic constraints \cite{AAD23_1, AAD23}. However, the most convenient for applications turn out to be the equations  written in a first-order (Hamiltonian) form (\ref{int01.1}). So, it is desirable to have a formalism that allows one to deduce these equations starting from the Lagrangian variational problem by direct application of the standard prescriptions of classical mechanics for the passage from Lagrangian to Hamiltonian formulations. The intermediate formalism seems to be the most economical way to do this. It should also be noted that a thorough analysis of the Lagrangian and Hamiltonian formulations reveals some specific properties of the formalism, which are not always taken into account in the literature, when formulating the laws of motion and applying them. In several cases this even leads to the need to revise some classical problems of the dynamics of a spinning body \cite{AAD23_8, AAD23_9}.

The remainder of the paper is organized as follows. In Sects. \ref{CAP1.2}  we shortly discuss the dynamics on a surface $G_\alpha(q^A)=0$ in terms of unconstrained variables, and outline the Liouville integration procedure in a form, convenient for the later comparison with the integration method based on Hamiltonian vector field. In Sect. \ref{CAP1.4} we describe Hamiltonian reduction on intermediate submanifold with use of Dirac bracket. In Sect. \ref{CAP1.5} we present our intermediate formalism for establishing the Poisson structure and Hamiltonian equations on the intermediate submanifold. In Sect. \ref{CAP1.6} we present the method of integration of a first-order equations with help of Hamiltonian vector field. In Sect. \ref{CAP1.6.1} we illustrate the intermediate formalism on a simple example of a point particle forced to move on a sphere.  In Sect. \ref{CAP1.7} we use the intermediate formalism to establish the Poisson structure that lie behind the Euler-Poisson equations of a spinning body, and write their general solution in terms of power series with respect to evolution parameter, and with the coefficients determined by derivatives of the Hamiltonian vector field.

\section{Motion on a surface in terms of unconstrained variables and Integrability according to Liouville.}\label{CAP1.2}
%Second-order (Lagrangian) equations for a system with kinematic constraints.

We assume that original Lagrangian is non-singular 
\begin{eqnarray}\label{int0}
\det\frac{\partial^2 L(q^A, \dot q^A)}{\partial\dot q^A\partial\dot q^B}\equiv\det M_{AB}\ne 0, 
\end{eqnarray}
and the particle $q^A$ was forced to move on $k$\,-dimensional surface ${\mathbb S}^k$ determined by $n-k$ functionally independent equations 
\begin{eqnarray}\label{int1}
G_\alpha(q^A)=0, \qquad \alpha= 1, 2, \ldots , n-k,  \qquad \mbox{rank}~ \frac{\partial G_\alpha}{\partial q^A}\equiv \mbox{rank} ~ G_{\alpha A}=k.
\end{eqnarray}
Let $x^i$, $i=1, 2, \ldots , k$ be local coordinates on ${\mathbb S}^k$, and $q^A(x^i)$ be parametric equations of ${\mathbb S}^k$: $G_\alpha(q^A(x^i))\equiv 0$ for any $x^i$. Then equations of motion follow from  the following unconstrained Lagrangian: 
\begin{eqnarray}\label{int01.00}
\tilde L(x^i, \dot x^i)\equiv L(q^A(x^i), \dot q^A(x^i))=L\left(q^A(x^i), \frac{\partial q^A}{\partial x^i}\dot x^i\right), 
\end{eqnarray}
and read as follows:
\begin{eqnarray}\label{int01.01}  
\frac{\partial\tilde L}{\partial x^i}-\frac{d}{dt}\frac{\partial\tilde L}{\partial\dot x^i}=0.
\end{eqnarray}
By construction, for any solution $x^i(t)$ to the problem (\ref{int01.01}), the trajectories  $q^A(x^i(t))$ lie on the surface (\ref{int1}). 
This recipe has clear justification \cite{Rub_1957,Arn_1,deriglazov2010classical} for the Lagrangians of the form $T=\frac12 m_A(\dot q^A)^2-U(q^A)$ with  $m_A>0$. For more general Lagrangians it should be taken as the definition of a particle forced to live on a surface. 

We add one more technical restriction, assuming that the matrix $M_{AB}$ is positive-definite, that 
is ${\bf Y}^TM{\bf Y}>0$ for any non zero column ${\bf Y}$. Then the matrix
\begin{eqnarray}\label{int01.02}
M_{ij}\equiv\frac{\partial\tilde L}{\partial\dot x^i\partial\dot x^j}=M_{AB}\frac{\partial q^A}{\partial x^i}\frac{\partial q^B}{\partial x^j}
\equiv(Q^T)_{iA}M_{AB}Q_{Bj},
\end{eqnarray}
is non degenerate, see Appendix. In view of this, for positive-definite $L(q^A, \dot q^A)$, the Lagrangian $\tilde L(x^i, \dot x^j)$ is non-singular.  

The Hamiltonian formulation in terms of unconstrained variables can be obtained as follows. Introduce the conjugate momenta $p_i=\partial\tilde L/\partial\dot x^i$ for $x^i$. As $\det M_{ij}\ne 0$, these equations can be resolved with respect to $\dot x^i$, say $\dot x^i=v^i(x^j, p_k)$. Using these equalities, we construct Hamiltonian by excluding $\dot x^i$ from the expression $H=p_i\dot x^i-\tilde L(x^i, \dot x^j)$. Then, with use of canonical Poisson brackets  $\{x^i, p_j\}=\delta^i_j$, the Hamiltonian equations of the theory are 
\begin{eqnarray}\label{int11}
\dot x^i=\{x^i, H\}=\frac{\partial H}{\partial p_i}, \qquad \dot p_i=\{p_i, H\}=-\frac{\partial H}{\partial q^i}. 
\end{eqnarray} 

If the Hamiltonian does not explicitly depend on time, it is an integral of motion. If, in addition, there are extra $k-1$ integrals of motion, then, according to the Liouville's theorem, a general solution to equations of motion can be found in quadratures (that is calculating integrals of some known functions and doing the algebraic operations). 

{\bf Liouville's theorem.}  Let the Hamiltonian equations (\ref{int11}) admit $k$ integrals of motion $F_1=H, F_2, F_3. \ldots , F_k$. We assume that they are in involution and functionally independent with respect to momenta 
\begin{eqnarray}
\{F_i, F_j \}=0,  \qquad \mbox{or} \quad \frac{\partial F_i}{\partial x^a}\frac{\partial F_j}{\partial p_a}=\frac{\partial F_j}{\partial x^a}\frac{\partial F_i}{\partial p_a}, \label{int14} \\
\det\frac{\partial F_i(x^j, p_k)}{\partial p_j}\ne 0. \qquad \qquad \quad \label{int15}
\end{eqnarray}
Then the equations of motion are integrable in quadratures. 

{\it Proof.} The proof consists in formulating a recipe for constructing the general solution. 

{\it (A)}. Consider the equations $F_i(x^i, p_j)=c_i=\mbox{const}$ for constant-level surface of integrals of motion.  
Due to the condition (\ref{int15}), they can be solved with respect to $p_i$
\begin{eqnarray}\label{int16}
F_i(x^i, p_j)=c_i, \quad \leftrightarrow \quad p_i=f_i(x^i, c_j), \quad \mbox{then} \quad F_i(x^p, f_j(x^n, c_k))=c_i. 
\end{eqnarray}
We first confirm that the vector function $f_i$ is a gradient of some scalar function. 
Omitting $x^j$, which we temporarily consider as parameters, we have $F_i(f_j(c_k))=c_i$, that is $F_i$ and $f_j$ are mutually inverse transformations. Calculating derivative of this equality with respect to $c_j$ we get
\begin{eqnarray}\label{int17}
\left.\frac{\partial F_i}{\partial p_a}\right|_{p=f}\frac{\partial f_a}{\partial c_j}=\delta_{ij}, \quad \mbox{then} \quad 
\frac{\partial f_i}{\partial c_a}\left.\frac{\partial F_a}{\partial p_j}\right|_{p=f}=\delta_{ij}, \quad \mbox{and} \quad \det\frac{\partial f_a}{\partial c_j}\ne 0. 
\end{eqnarray} 
Contracting  Eq. (\ref{int14}) with $\partial f_k/\partial c_i$ and using (\ref{int17})  we get the identity
\begin{eqnarray}\label{int18}
\left.\frac{\partial F_j}{\partial x^k}\right|_{p=f}=\left.\frac{\partial f_k}{\partial c_b}\frac{\partial F_b}{\partial x^a}\frac{\partial F_j}{\partial p_a}\right|_{p=f}.
\end{eqnarray} 
Contracting $\partial f_b/\partial c_j$ with derivative of (\ref{int16}) with respect to $x^k$ and using Eq. (\ref{int17}) we get the following expression for the derivative of $f_b$
\begin{eqnarray}\label{int19}
\frac{\partial f_b}{\partial c_j}\left[\left.\frac{\partial F_j}{\partial x^k}\right|_{p=f}+
\left.\frac{\partial F_j}{\partial p_a}\right|_{p=f}\frac{\partial f_a}{\partial x^k}\right]=0, \quad \mbox{then} \quad 
\frac{\partial f_b}{\partial x^k}=-\frac{\partial f_b}{\partial c_j}\left.\frac{\partial F_j}{\partial x^k}\right|_{p=f}. 
\end{eqnarray} 
Together with (\ref{int18}) this implies that $\partial_k f_b$ is antisymmetric matrix
\begin{eqnarray}\label{int20}
\frac{\partial f_b}{\partial x^k}=-\frac{\partial f_b}{\partial c_j}\left.\frac{\partial F_j}{\partial x^k}\right|_{p=f}=- 
\frac{\partial f_b}{\partial c_j}\left.\frac{\partial f_k}{\partial c_n}\frac{\partial F_n}{\partial x^a}\frac{\partial F_j}{\partial p_a}\right|_{p=f}=-
\frac{\partial f_k}{\partial c_n}\left.\frac{\partial F_n}{\partial x^b}\right|_{p=f}=\frac{\partial f_k}{\partial x^b}, \quad \mbox{or} \quad \partial_if_j-\partial_jf_i=0. 
\end{eqnarray}
Then Eqs. (\ref{int14}) imply that the quantities $p_i-f_i(x^k, c_j)$ are in involution
\begin{eqnarray}\label{int22}
\{p_i-f_i, p_j-f_j\}=0.
\end{eqnarray} 
According to (\ref{int20}),  $f_i(x^k)$ is a curl-free vector field. So there is the potential $\Phi$: $f_i(x^k, c_j)=\partial\Phi(x^k, c_j)/\partial x^i$.  In the result, we demonstrated that equations of constant-level surface (\ref{int16})  can be written in the form 
\begin{eqnarray}\label{int21}
p_i=\partial_i\Phi(x^j, c_k).
\end{eqnarray}

{\it (B)}. According to Stokes' theorem, the line integral of a curl-free field does not depend on the choice of the integration path, and gives the potential 
\begin{eqnarray}\label{int23}
\Phi(x^k, c_j)=\int_{0}^{x^k} f_i(z^i, c_j) dz^i. 
\end{eqnarray} 

{\it (C)}. Substituting the solution (\ref{int21}) to the equation $H(x^i, p_j)=c_1\equiv E$ into this equation, we have the identity
\begin{eqnarray}\label{int24}
H\left(x^i, \frac{\partial\Phi(x^i, c_j)}{\partial x^j}\right)=E.
\end{eqnarray} 
Then the function 
\begin{eqnarray}\label{int25}
S(t, x^i, c_j)=-Et+\Phi(x^i, c_j), 
\end{eqnarray} 
with the property
\begin{eqnarray}\label{int26}
\det\frac{\partial^2 S}{\partial x^i\partial c_j}=\det\frac{\partial f_i}{\partial c_j}\ne 0,
\end{eqnarray} 
by construction obeys the Hamilton-Jacobi equation
\begin{eqnarray}\label{int27}
\frac{\partial S}{\partial t}+H\left(x^i, \frac{\partial S}{\partial x^j}\right)=0. 
\end{eqnarray} 
According to the theory of canonical transformations (see Sect. 4.7 in \cite{deriglazov2010classical}), the general solution to the Hamiltonian equations (\ref{int11}) with $2k$ integration constants $c_k, b_i$ can be now obtained solving the algebraic equations 
\begin{eqnarray}\label{int28}
p_i=\frac{\partial S(t, x^j, c_k)}{\partial x^i}=f_i(x^j, c_k), \qquad b_i=\frac{\partial S(t, x^j, c_k)}{\partial c_i}=-t\delta^i{}_E+\frac{\partial\Phi(x^j, c_k)}{\partial c_i}.
\end{eqnarray} 
with respect to $x^i$ and $p_j$. The resolvability of the second equation is guaranteed by (\ref{int26}). 

As a result, the problem of integrating the Hamiltonian system (\ref{int11}) is reduced to the calculation of line integral (\ref{int23}). In turn, this can be reduced to calculation of definite integrals. To see this, let us specify the equations (\ref{int28}) to the case of a theory with two configuration-space variables $x^i=(x, y)$ and two integrals of motion $H(x, y, p_x, p_y)=E$ and $F(x, y, p_x, p_y)=c$. Solving these algebraic equations we get $p_x=f_x(x, y, E, c)$ and $p_y=f_y(x, y, E, c)$. Taking the path of integration to be the pair of intervals, $(0, 0)\rightarrow (x, 0)\rightarrow (x, y)$, we obtain the potential 
\begin{eqnarray}\label{int29}
\Phi(x, y, E, c)=\int_{0}^{x} f_x(x', 0, E, c)dx'+\int_{0}^{y} f_y(x, y', E, c)dy'. 
\end{eqnarray} 
Then Eqs. (\ref{int28}) read as follows
\begin{eqnarray}\label{int30}
p_x=f_x(x, y, E, c), \qquad p_y=f_y(x, y, E, c), \qquad \qquad \qquad \quad \cr 
b_x=-t+\int_{0}^{x} \frac{\partial f_x(x', 0, E, c)}{\partial E}dx'+\int_{0}^{y} \frac{\partial f_y(x, y', E, c)}{\partial E}dy', \cr
b_y=\int_{0}^{x} \frac{\partial f_x(x', 0, E, c)}{\partial c}dx'+\int_{0}^{y} \frac{\partial f_y(x, y', E, c)}{\partial c}dy'. \qquad ~  
\end{eqnarray} 
So the problem is reduced to the calculation of four definite integrals indicated in these equations.

\section{Motion on a surface in terms of original variables.}\label{CAP1.4}

To work with a particle on a surface in terms of original variables, we can use the variational problem with the modified Lagrangian (\ref{int02}), where the constraints are taken into account with help of auxiliary dynamical variables $\lambda_\alpha(t)$, called Lagrangian multipliers. In all calculations they should be treated on equal footing with $q^A(t)$. In particular, looking for the equations of motion, we take variations with respect to $q^A$ and  
all $\lambda_\alpha$.
The variation with respect to $\lambda_\alpha$, implies 
\begin{eqnarray}\label{int2}
G_\alpha(q^A)=0, 
\end{eqnarray}
that is the constraints arise as a part of conditions of extreme of the action functional. So the presence of $\lambda_\alpha$ allows $q^A$ to be treated as unconstrained variables, that should be varied independently in obtaining the equations of motion.  Taking the variation with respect to $q^A$ we get
\begin{eqnarray}\label{int3}
-\frac{d}{dt}\frac{\partial L}{\partial\dot q^A}+\frac{\partial L}{\partial q^A}-\lambda_\alpha G_{\alpha A}=0.  
\end{eqnarray}
Computing the time derivative, these equations read
\begin{eqnarray}\label{int4}
\ddot q^A=K^A(q, \dot q)-\tilde M^{AB}G_{\beta B}\lambda_\beta, 
\end{eqnarray}
where $\tilde M^{AB}$ is the inverse of $M_{AB}(q^A, \dot q^B)$ and $K^A\equiv\tilde M^{AB}\left[-\dot q^C\partial^2 L/(\partial\dot q^B\partial q^C)+
\partial L/\partial q^B\right]$.  The theories (\ref{int4}) and (\ref{int01.01}) turn out to be equivalent, see \cite{Arn_1,deriglazov2010classical}.

The auxiliary variables $\lambda_\alpha$ can be excluded from the system (\ref{int3})  (or (\ref{int4})) as follows. For any solution $q^A(t)$, the identity $G_\alpha(q^A(t))=0$ implies $\dot G_\alpha=G_{\alpha A}\dot q^A=0$. Calculating one more derivative we get $G_{\alpha A}\ddot q^A+\partial_B\partial_A G_\alpha\dot q^A\dot q^B=0$. Using expression for $\ddot q^A$ from (\ref{int4}) we get
\begin{eqnarray}\label{int5}
C_{\alpha\beta}\lambda_\beta=G_{\alpha A}K^A+\partial_A\partial_B G_\alpha \dot q^A\dot q^B, 
\quad \mbox{where} \quad 
C_{\alpha\beta}\equiv G_{\alpha A}\tilde M^{AB}G_{\beta B}.
\end{eqnarray}
According to Appendix, $C$ has the inverse matrix $\tilde C$, so we can separate $\lambda_\beta$ as follows
\begin{eqnarray}\label{int6}
\lambda_\beta=\tilde C_{\beta\alpha}[G_{\alpha A}K^A+\partial_A\partial_B G_\alpha\dot q^A\dot q^B]. 
\end{eqnarray}
Inserting this $\lambda_\beta$ into Eq. (\ref{int3}) or (\ref{int4}), we obtain closed equations for determining the physical variables $q^A(t)$. 

{\bf Comment.}  If the condition (\ref{int01.02}) is not satisfied, the invertibility of $C$ is not guaranteed, and we need to continue the analysis of the system (\ref{int5}). The general procedure can be found in the Appendix C of \cite{GT}. Here we will only show that in a theory with kinematic constraints the auxiliary variables can always be excluded from the equations for physical variables. Without loss of generality, we can assume that the coordinates $q^A$ were enumerated in such a way that non-vanishing minor of the matrix $G_{\alpha A}$ is located in  the first $n-k$ columns, then  
\begin{eqnarray}\label{int7}
q^A=(q^\alpha, q^i), \qquad \det\frac{\partial G_\alpha}{\partial q^\beta}\ne 0.
\end{eqnarray}
Let us consider the original theory (\ref{int02}) in special coordinates $q'^A$, adapted to the surface and defined as follows: 
\begin{eqnarray}\label{int8}
q^A=(q^\alpha, q^i) ~ \leftrightarrow ~ q'^A=(q'^\alpha=G_{\alpha}(q^A), ~ q'^i=q^i). 
\end{eqnarray}
That is we taken the constraint's functions $G_\alpha(q^A)$ as a part of new coordinates. 
In the adapted coordinates our surface is just the hyperplane $q'^\alpha=0$, and $q'^i$ can be taken as its local coordinates.  For the inverse transformation we get
\begin{eqnarray}\label{int9}
q^A=(q^\alpha=\tilde G_\alpha(q'^A), ~ q^i=q'^i) \quad \mbox{then} \quad \dot q'^\alpha=\frac{\partial\tilde G_\alpha}{\partial q'^A}\dot q'^A, \quad \dot q^i=\dot q'^i, 
\end{eqnarray}
where $\tilde G_\alpha(q'^A)$ is the solution to equations $q'^\alpha=G_\alpha(q^\alpha, q'^i)$: $G_\alpha(\tilde G_\beta(q'^A), q'^i))=q'^\alpha$. An invertible change of variables can be made directly in the Lagrangian (\ref{int02}), this leads to an equivalent formulation of the original theory, see Sect. 1.4.2 in \cite{deriglazov2010classical}. Substituting the expressions (\ref{int9}) into (\ref{int02}) we get the Lagrangian of the form $L'(q'^A, \dot q'^A)-\lambda_\alpha q'^\alpha$, which implies equations of the following structure: 
\begin{eqnarray}
\lambda_\alpha=A_\alpha(q'^A, \dot q'^A, \ddot q'^A), \qquad \qquad  \label{int10} \\ 
B_i(q'^A, \dot q'^A, \ddot q'^A)=0, \qquad q'^\alpha=0. \label{int11.1}
\end{eqnarray}
That is we have closed system (\ref{int11.1}) for determining $q'^A(t)$, while $\lambda_\alpha(t)$ then can be found algebraically from (\ref{int10}). 

For the latter use, observe that 
\begin{eqnarray}\label{int11.11}
M'_{AB}=\frac{\partial^2 L'}{\partial\dot q'^A\partial\dot q'^B},
\end{eqnarray}
and its inverse are positive-definite matrices together with $M_{AB}$.

{\bf Hamiltonian formulation of the theory (\ref{int02}) on phase space $(q^A, p_B)$.} Without loss of generality,  we assume that equations of the surface $G_\alpha(q^A)=0$  can be resolved with respect to first $n-k$\,-coordinates. In accordance to this, the set $q^A$ is divided on two subgroups, $q^\alpha$ and $q^i$. Greek indices from the beginning of the alphabet run from $1$ to $n-k$, while Latin indices from the middle of the alphabet run from $1$ to $k$. So 
\begin{eqnarray}\label{p3.1}
{\mathbb S}^k=\{q^A=(q^\alpha , q^i),  ~ G_\alpha(q^A)=0, ~ \det\left.\frac{\partial G_\alpha}{\partial q^\beta }\right|_{\mathbb S}=n-k, ~ \alpha=1, 2, ~ \ldots, n-k\}, 
\end{eqnarray}
and our variational problem is  (\ref{int02}). 
Applying the Dirac's method, we introduce conjugate momenta $p_A=\partial L/\partial\dot q^A$ and $p_{\lambda\alpha}=\partial L/\partial\dot\lambda_\alpha$ for all configuration-space variables $q^A$ and $\lambda_\alpha$. Conjugate momenta 
for $\lambda_\alpha$ are the primary constraints: $p_{\lambda\alpha}=0$. Since the Lagrangian $L$ was assumed non-singular, the expressions for $p_A$ can be resolved with respect to velocities: 
\begin{eqnarray}\label{p3.2}
p_A=\frac{\partial L}{\partial\dot q^A}, \qquad \mbox{then} \quad  \dot q^A=v^A(q, p).
\end{eqnarray} 
To find the Hamiltonian, we exclude velocities from the expression $H=p_A\dot q^A-(L-\lambda_\alpha G_\alpha)+
\varphi_\alpha p_{\lambda_\alpha}$, obtaining 
\begin{eqnarray}\label{p3.3}
H=H_0+\lambda_\alpha G_\alpha(q^A)+\varphi_\alpha p_{\lambda_\alpha}, \quad \mbox{where} 
\quad  H_0\equiv p_A v^A(q, p)-L(q^A, v^B(q, p)). 
\end{eqnarray}
By $\varphi_\alpha $ we denoted the Lagrangian multipliers for the primary constraints.  
Preservation in time of the primary constraints, $\dot p_{\lambda_\alpha}=\{p_{\lambda_\alpha}, H\}=0$ implies $G_\alpha=0$ as the secondary constraints. In turn, the equation $dG_\alpha/dt=\{G_\alpha, H\}=\{G_\alpha, H_0\}=0$ implies tertiary constraints, that should be satisfied by all true solutions 
\begin{eqnarray}\label{p6}
\Phi_{\alpha}\equiv \{G_\alpha, H_0\}= G_{\alpha B}(q)v^B(q, p)=0, \qquad \mbox{where} \quad G_{\alpha B}\equiv\frac{\partial G_\alpha(q)}{\partial q^B}. 
\end{eqnarray}
The Lagrangian counterpart of these constraints is $\dot q^A\partial_A G_\alpha=0$, and mean that for true trajectories the velocity vector is 
tangent to the surface ${\mathbb S}^k$. 
Calculate
\begin{eqnarray}\label{p7}
\mbox{rank}~\frac{\partial\Phi_{\alpha}}{\partial p_B}=\mbox{rank}~(G_{\alpha A}\tilde M^{AB})=n-k, \qquad \mbox{where} \quad  
\tilde M^{AB}\equiv\frac{\partial v^A(q, p)}{\partial p_B}. 
\end{eqnarray}
Note that $\tilde M^{AB}$ is inverse of the Hessian matrix $M_{AB}$. 
This implies that the constraints $\Phi_\alpha$ are functionally independent and can be resolved with respect to some $n-k$ momenta of the set $p_A$. This implies also that the constraints  $G_\beta$ and $\Phi_\alpha$ are functionally independent. Calculating their Poisson brackets we get the matrix
\begin{eqnarray}\label{p8.0}
\{ G_\alpha, \Phi_\beta \}=G_{\beta A}(q)\tilde M^{AB} G_{\alpha B}\equiv b_{\alpha\beta}.
\end{eqnarray}
For our Lagrangian with positive-definite $M_{AB}$, this matrix is non degenerate, see Appendix.  

For the latter use, we introduce the matrix composed by brackets of the constraints $T_I=(G_\alpha, \Phi_\beta)$
\begin{eqnarray}\label{p8.1}
\triangle_{IJ}=\left(
\begin{array}{cc}
0 & b \\
-b^T& c 
\end{array}\right), \qquad 
\triangle^{-1}_{IJ}=\left(
\begin{array}{cc}
b^{-1 T}cb^{-1} & -b^{-1 T} \\
b^{-1} & 0 
\end{array}\right).
\end{eqnarray}
where the first block corresponds to $\{G_\alpha, G_\beta\}=0$, and $c_{\alpha\beta}=\{\Phi_\alpha, \Phi_\beta\}$. As $b$ is invertible, the matrix $\triangle_{IJ}$ is invertible, so our constraints $G_\beta$ and $\Phi_\alpha$ are of second class. 

Preservation in time of the tertiary constraints gives fourth-stage constraints that involve $\lambda_\alpha$, and can be used to find them through $q^A$ and $p_B$
\begin{eqnarray}\label{p8.2}
\lambda_\alpha=b^{-1 T}_{\alpha\beta}\{\Phi_\beta, H_0\}.
\end{eqnarray}
At last, preservation in time of the fourth-stage constraints gives an equation that algebraically determines the Lagrangian 
multipliers $\varphi_\alpha$ through other variables 
\begin{eqnarray}\label{p8.3}
\varphi_\alpha=b^{-1}_{\alpha\beta}\left[\{\{\Phi_\alpha, H_0\}, H_0\}+\lambda_\gamma\{\{\Phi_\alpha, H_0\}, G_\gamma\}-
\{b_{\beta\gamma}, H_0\}\lambda_\gamma-\{b_{\beta\gamma}, G_\sigma\}\lambda_\gamma\lambda_\sigma\right]. 
\end{eqnarray}
In the absence of new constraints, the Dirac's procedure is over. 

In resume, we revealed the following chain of constraints
\begin{eqnarray}\label{p8.4}
p_{\lambda\alpha}=0, \quad G_\alpha(q)=0, \quad \Phi_\alpha\equiv G_{\alpha B}(q)v^B(q, p)=0, \quad \lambda_\alpha=b^{-1 T}_{\alpha\beta}\{\Phi_\beta, H_0\},
\end{eqnarray}
and determined the auxiliary variables $\varphi_\alpha$. Note that the phase-space variable $p_{\lambda\alpha}$ is just a constant, while $\lambda_\alpha$ is presented through $q^A$ and $p_B$. So we only need to write the dynamical equations for $q^A$ and $p_B$. 
The variables $\lambda_\alpha$ can be excluded from the Hamiltonian (\ref{p3.3}) using the constraint (\ref{p8.2}). Besides, we can omit the term $\varphi_\alpha p_{\lambda\alpha}$, since it does not contribute into Hamiltonian equations for the phase-space  variables $q^A$, $p_B$. With the resulting Hamiltonian, the equations read as follows:
\begin{eqnarray}\label{p8.5}
\dot q^A=\{q^A, H_0+b^{-1 T}_{\alpha\beta}\{\Phi_\beta, H_0\}G_\alpha\}=\{q^A, H_0\}+\{q^A, G_\alpha\}b^{-1 T}_{\alpha\beta}\{\Phi_\beta, H_0\}, \cr
\dot p_A=\{q_A, H_0+b^{-1 T}_{\alpha\beta}\{\Phi_\beta, H_0\}G_\alpha\}=\{p_A, H_0\}+\{p_A, G_\alpha\}b^{-1 T}_{\alpha\beta}\{\Phi_\beta, H_0\}. 
\end{eqnarray}
Writing the last equalities we taken into account that $G_\alpha=0$ for true solutions. 

Dirac noticed, that these equations can be rewritten in terms of canonical Hamiltonian without auxiliary variables
\begin{eqnarray}\label{p8.5.1}
H_0(q^A, p_B)=p_Av^A(q, p)-L(q^A, v^B(q, p)),  
\end{eqnarray}
if instead of canonical Poisson bracket we introduce the famous Dirac bracket. Given two phase-space functions $A(q, p)$ and $B(q,p)$, their Dirac bracket is 
\begin{eqnarray}\label{p8.7}
\{A, B\}_D=\{A, B\}-\{A, T_I\}\triangle^{-1}_{IJ}\{T_J, B\}. 
\end{eqnarray}
This has all properties of the canonical Poisson bracket, including antisymmetry and the Jacobi identity \cite{AAD_2022}. Besides, its remarkable property is that $T_I=(G_\alpha, \Phi_\beta)$ represent its Casimir functions, that is Dirac bracket of any phase-space function with any constraint $T_I$ vanishes: $\{A, T_I\}_D=0$. 
The equations constructed with help of $H_0$ and Dirac bracket
\begin{eqnarray}\label{p8.6}
\dot q^A=\{q^A, H_0\}_D, \qquad \dot p_A=\{p_A, H_0\}_D, 
\end{eqnarray}
differ from (\ref{p8.5}) by terms proportional to the constraints, and therefore are equivalent.  The final equations (\ref{p8.6}) do not involve the auxiliary variables and are written on the phase space $(q^A, p_B)$. The Dirac bracket determines the Poisson structure of this space. 

{\bf Hamiltonian reduction to the intermediate submanifold.} Using the Dirac formalism, we obtained $2n+ 2(n-k)$ equations of our theory written on $2n$\,-dimensional phase space with coordinates $(q^A, p_B)$. They are the dynamical equations (\ref{p8.6}) and the constraints $G_\alpha(q^A)=0$ and $\Phi_\alpha(q^A, p_B)=0$. All solutions to our equations live on $2k$\,-dimensional submanifold specified by these algebraic constraints. They could be used to exclude $2(n-k)$ variables from the formalism. However, as we saw above, it may be desirable to work with our theory keeping all $q^A$. Therefore we exclude only a part of momenta, making reduction of our theory to intermediate submanifold of equations $\Phi_\alpha(q^A, p_B)=0$.  Let 
\begin{eqnarray}\label{p8.10}
p_\alpha=f_\alpha(q^A, p_i). 
\end{eqnarray}
be a solution to the constraints $\Phi_\alpha(q^A, p_B)=0$. 
The reduction can be done while at the same time keeping the Hamiltonian character of resulting equations, that is we establish Poisson structure and  Hamiltonian for our equations on the intermediate submanifold with the coordinates $(q^A, p_i)$.  Because of the property that the constraints are composed of Casimir functions, the reduction consists in elimination the variables $p_\alpha$ from the formalism as follows. \par

\noindent {\bf 1.}  It is known \cite{AAD_2022} that together with $\Phi_\alpha=0$, the functions $p_\alpha-f_\alpha(q^A, p_i)$ also represent Casimir functions of the Dirac bracket, so for any phase-space function $A(q^A, p_B)$ we get 
\begin{eqnarray}\label{p8.11}
\{A(q^A, p_B), p_\alpha\}_D=\{A(q^A, p_B), f_\alpha(q^A, p_i)\}_D. 
\end{eqnarray}
As a consequence, computation of the Dirac bracket and substitution (\ref{p8.10}) are commuting operations:
\begin{eqnarray}\label{p8.12}
\{A(q^A, p_B), B(q^A, p_B)\}_D=\{A(q^A, p_i, f_\alpha), B(q^A, p_i, f_\alpha)\}_D. 
\end{eqnarray} \par
\noindent {\bf 2.} Using (\ref{p8.7}) and (\ref{p8.10}), we define the following brackets on the submanifold $(q^A, p_i)$: 
\begin{eqnarray}\label{p8.13}
\{A(q^A, p_i), B(q^A, p_i)\}'=\left.\{A(q^A, p_i), B(q^A, p_i)\}_D\right|_{p_\alpha=f_\alpha(q^A, p_i)}. 
\end{eqnarray}
Because of the property (\ref{p8.11}) the brackets $\{ {}, {} \}'$ obey the Jacobi identity (for the direct proof, see Sect. 4.2 in \cite{AAD_2022}), and hence determine the Poisson structure on the submanifold $(q^A, p_i )$. \par

\noindent {\bf 3.}  Let us replace $p_\alpha$ on $f_\alpha(q^A, p_i)$ in the Hamiltonian (\ref{p8.5.1}), denoting the resulting expression by $H'_0(q^A, p_j)$  
\begin{eqnarray}\label{p8.14}
H'_0(q^A, p_j)=\left.\left[p_Av^A(q, p)-L(q^A, v^B(q, p))\right]\right|_{p_\alpha=f_\alpha(q^A, p_i)}. 
\end{eqnarray}
Because of the property (\ref{p8.11}), $H_0$ can be used in Eqs. (\ref{p8.6}) instead of $H$, this will give an equivalent Hamiltonian equations. Replacing $p_\alpha$ according to (\ref{p8.10}) in the r.h.s. of these equations, we get an equivalent equations with the bracket (\ref{p8.13}) 
\begin{eqnarray}\label{p8.15}
\dot q^A=\{q^A, H'_0( q^B, p_j) \}'  \qquad \dot p_i=\{p_i, H'_0( q^B, p_j) \}'. 
\end{eqnarray}
Together with the algebraic equations $G_\alpha=0$ and $p_\alpha=f_\alpha(q^A, p_i)$ they are equivalent to the original 
system composed of (\ref{p8.6}), $G_\alpha=0$ and $\Phi_\alpha=0$.  This completes the procedure of reduction to the intermediate submanifold $\Phi_\alpha=0$.

\section{Intermediate formalism.}\label{CAP1.5}  

Here we present more economic way to construct Hamiltonian formulation of the theory (\ref{int02}) on the intermediate submanifold, which does not require constructing the Dirac bracket  and then reducing it to the submanifold. 

To this aim we rewrite the obtained Hamiltonian theory (\ref{p8.4}), (\ref{p8.5}) in non-canonical phase-space coordinates  with special properties. The matrix $G_{\alpha B}(q^A)$ of Eq. (\ref{p6}) is composed by $(n-k)$ linearly independent vector fields ${\bf G}_\alpha(q^A)$ orthogonal to the surface ${\mathbb S}^k$ of the configuration space $q^A$. Let us consider the linear 
system $G_{\alpha B} x_B=0$. It  has a general solution\footnote{To avoid a possible confusion, we point out that in the similar equation (\ref{p6}), representing the tertiary constraints, $f^A$ are given functions of $q$ and $p$.} of the form $x_B=c^i G_{i B}$, where the linearly independent vectors ${\bf G}_i$ are fundamental solutions to this system. They have the following structure: 
\begin{eqnarray}\label{p9}
{\bf G}_i=( ~G_{i1}(q), G_{i2}(q), \ldots , G_{i, n-k}(q), 0, \ldots , 1, 0 \ldots , 0 ~ ), \qquad \mbox{then} \quad G_{\alpha B}G_{i B}=0. 
\end{eqnarray}
By construction, these vector fields form a basis of tangent space to the surface ${\mathbb S}^k$. Together with ${\bf G}_\alpha$, they form a basis of tangent space to the entire configuration space. Using the rows ${\bf G}_\beta$ and ${\bf G}_j$, we construct an invertible matrix $G_{BA}$, and use it to define the new momenta $\pi_B$ of the phase space $(q^A, p_B)$ as follows: 
\begin{eqnarray}\label{p10}
G_{BA}(q)=\left(
\begin{array}{c}
G_{\beta A} \\
G_{j A}
\end{array}
\right), \qquad \pi_B=G_{BA}(q) p_A, \quad \mbox{then} \quad p_A=G^{-1}_{AB}(q)\pi_B\equiv\tilde G_{AB}(q)\pi_B. 
\end{eqnarray}
Let us take $q^A$ and $\pi_B$ as the new phase-space coordinates. 
Their special property is that both $q^A$ and $\pi_i$ have vanishing brackets with the original constraints $G_\alpha$
\begin{eqnarray}\label{p11}
\{ q^A, G_\alpha \}=0, \qquad \{\pi_i, G_\alpha \}=0, 
\end{eqnarray}
the latter equality is due to Eq. (\ref{p9}). 

Let us rewrite our theory in the new variables.  Using the canonical brackets $\{q^A, p_B\}=\delta^A{}_B$, we get Poisson brackets of the new variables
\begin{eqnarray}\label{p12}
\{ q^A, q^B\}=0, \qquad \{q^A, \pi_B \}=G_{BA}(q), \qquad \{\pi_A, \pi_B \}=-c_{AB}{}^D(q)\tilde G_{DE}(q)\pi_E, 
\end{eqnarray}
where the Lie brackets of basic vector fields ${\bf G}_A$ appear
\begin{eqnarray}
c_{AB}{}^D=[{\bf G}_A, {\bf G}_B]^D\equiv G_{AE}\partial_E G_{BD}-G_{BE}\partial_E G_{AD}, \label{p13}\\
c_{ij}{}^k=0. \qquad \qquad \qquad \qquad \qquad \qquad \label{p13.1}
\end{eqnarray} 
Therefore the Lie bracket of the vector fields ${\bf G}_A$ determines Poisson structure of our theory in the sector $\pi_A$. 
The structure functions $c_{ij}{}^k$ vanish for our choice of basic vectors ${\bf G}_i$ of special form, see Eq. (\ref{p9}).  In particular, the
Poisson brackets of the coordinates $q^A$ and $\pi_i$ are  
\begin{eqnarray}\label{p12.1}
\{ q^A, q^B\}=0, \qquad \{q^A, \pi_i \}=G_{iA}(q), \qquad \{\pi_i, \pi_j \}=-c_{ij}{}^\alpha(q)\tilde G_{\alpha E}(q)\pi_E.  
\end{eqnarray}
The Hamiltonian (\ref{p3.3}) reads
\begin{eqnarray}\label{p14}
H=H_0+\lambda_\alpha G_\alpha(q), \qquad H_0=\tilde G_{AC}\pi_C v^A(q, \tilde G\pi)-L(q^A, v^B(q, \tilde G\pi)). 
\end{eqnarray}
At last, our second-class constraints in the new coordinates are
\begin{eqnarray}\label{p15}
G_\alpha (q)=0, \qquad \Phi_\alpha\equiv G_{\alpha A}(q)v^A(q, \tilde G\pi)=0.
\end{eqnarray}

Let us confirm that the tertiary constraints $\Phi_\alpha$  can be resolved with respect to $\pi_\alpha$. To this aim we compute the 
matrix $\partial\Phi_\alpha/\partial\pi_\beta$, and show that its determinant is not zero
\begin{eqnarray}\label{p19}
\det\frac{\partial(G_{\alpha A}v^A(q, \tilde G\pi))}{\partial\pi_\beta}=\det [G_{\alpha A}\tilde M^{AD}(q, \tilde G\pi)\tilde G_{D \beta}]\ne 0.
\end{eqnarray}
It is not zero for our class of positive-definite Lagrangians (\ref{int01.02}), see Appendix. Resolving the constraints $\Phi_\alpha=0$, say  
\begin{eqnarray}\label{p19.0}
\pi_\alpha=f_\alpha(q^A, \pi_i), 
\end{eqnarray}
we use the resulting expressions to exclude $\pi_\alpha$ from (\ref{p12.1}) and (\ref{p14}), thus obtaining 
\begin{eqnarray}\label{p20}
\{ q^A, q^B \}'=0, \qquad \{q^\alpha, \pi_i \}'=G_{i \alpha}(q), \qquad \{q^j, \pi_i \}'=\delta^j{}_i,  \qquad 
\{\pi_i, \pi_j \}'=-c_{ij}{}^\alpha[\tilde G_{\alpha k}\pi_k+\tilde G_{\alpha\beta}f_\beta(q^A, \pi_i)]. 
\end{eqnarray}
\begin{eqnarray}\label{p14.1}
H'_0(q^A, \pi_j)=\left.\tilde G_{AC}\pi_C v^A(q, \tilde G\pi)-L(q^A, v^B(q, \tilde G\pi))\right|_{\pi_\alpha=f_\alpha(q^A, \pi_i)}. 
\end{eqnarray}
In general, the brackets (\ref{p20}) are non-linear for both $q^A$ and $\pi_i$. Their dependence on the choice of tangent vector fields ${\bf G}_i$ to the surface ${\mathbb S}^k$ is encoded in three places: in the  brackets $\{q^\alpha, \pi_i\}'$, in the matrix $\tilde G$, as well as in the structure functions $c_{ij}{}^\alpha$, see Eq. (\ref{p13}). 

Using these brackets and Hamiltonian, let us write the following system of equations:
\begin{eqnarray}
\dot q^A=\{q^A, H'_0( q^B, \pi_j) \}', \qquad \dot\pi_i=\{\pi_i, H'_0( q^B, \pi_j) \}'; \label{p14.2} \\
G_\alpha(q)=0, \qquad \pi_\alpha=f_\alpha(q^A, \pi_i). \label{p14.3} \qquad \quad \qquad 
\end{eqnarray}

{\bf Affirmation.} The brackets (\ref{p20}) obey the Jacobi identity and hence determine the Poisson structure on the intermediate submanifold $\Phi_\alpha=0$ equipped with the coordinates $(q^A, \pi_i)$. Besides,  equations (\ref{p14.2}) and (\ref{p14.3}) represent an equivalent formulation of the original theory (\ref{p8.6}), (\ref{p8.4}). \par

\noindent {\it Proof.} To establish the equivalence, we consider our theory in the variables $(q^A, \pi_B)$, write the Dirac bracket in these variables and then reduce it to the intermediate submanifold. 

Using the constraints (\ref{p15}), we construct Dirac bracket on the phase space $(q^A, \pi_B)$ as follows: 
\begin{eqnarray}\label{p16}
\{ A, B\}_D=\{A, B\}-\{A, T_I\}\triangle^{-1}_{IJ}\{T_J, B\}.
\end{eqnarray}
Here $T_I$ is the set of all constraints: $T_I=(G_\alpha(q), \Phi_\beta(q, \pi))$. Besides, denoting symbolically the blocks $b=\{ G, \Phi \}$ 
and $c=\{ \Phi, \Phi  \}$, the matrices $\triangle$ and $\triangle^{-1}$ are   
\begin{eqnarray}\label{p17}
\triangle=\left(
\begin{array}{cc}
0 & b \\
-b^T& c 
\end{array}\right), \qquad 
\triangle^{-1}=\left(
\begin{array}{cc}
b^{-1 T}cb^{-1} & -b^{-1 T} \\
b^{-1} & 0 
\end{array}\right).
\end{eqnarray}
The constraint's functions (\ref{p15}) are Casimir functions of the Dirac bracket (\ref{p16}). 
Similarly to the previous section, as Hamiltonian equations of our theory we can take  
\begin{eqnarray}\label{p17.1}
\dot q^A=\{q^A, H\}_D, \qquad \dot\pi_A=\{\pi_A, H\}_D, 
\end{eqnarray}
with $H$ written in Eq. (\ref{p14}).
Eq. (\ref{p17}) implies the following structure of the Dirac bracket
\begin{eqnarray}\label{p18}
\{ A, B\}_D=\{A, B\}-\{A, G\}\triangle ' \{G, B\}+\{A, G\}\triangle '' \{\Phi, B\}, 
\end{eqnarray}
that is the last two terms on r.h.s. involve at least one constraint $G_\alpha$. 
Taking into account Eqs. (\ref{p11}), we conclude that in the passage from Poisson bracket (\ref{p12}) to the Dirac bracket (\ref{p16}), the brackets (\ref{p12.1}) of basic variables $q^A$ and $\pi_i$ will not be modified, retaining their original form. Excluding $\pi_\alpha$ from their right hand sides with help of (\ref{p19.0}), we arrive at the brackets (\ref{p20}). Since $\pi_\alpha-f_\alpha(q^A, \pi_i)$ are Casimir functions of the Dirac bracket (\ref{p16}), the brackets (\ref{p20}) obey the Jacobi identity, see Sect. 4.2 in \cite{AAD_2022} for the direct proof. 

To reduce the equations (\ref{p17.1}) to the intermediate submanifold $\Phi_\alpha=0$, we proceed in the same way as in the previous section. First, working with Eqs. (\ref{p17.1}), we can omit the terms with constraints in the Hamiltonian (\ref{p14}), and then use (\ref{p19.0}) in the resulting expression. This gives the Hamiltonian (\ref{p14.1}), which therefore can be used instead of $H$ in equations (\ref{p17.1}) for $q^A$ and $\pi_i$. Second, excluding $\pi_\alpha$ from r.h.s. of these equations with help of (\ref{p19.0}), they acquire the form (\ref{p14.2}). This completes the proof of the affirmation.

{\bf Another set of non-canonical variables.} Instead of (\ref{p10}), we can equally consider the following non-canonical set $q^A$, $\pi_B$: 
\begin{eqnarray}\label{p21}
\pi_\alpha=G_{\alpha A}v^A(q, p)\equiv\Phi_\alpha, \qquad \pi_i=G_{i A}p_A.  
\end{eqnarray}
That is we taken the third-stage constraints $\Phi_\alpha$ as a part of new momenta. 
Using the adapted coordinates (\ref{int8}), we conclude that the change (\ref{p21}) is invertible with respect to $p_A$
\begin{eqnarray}\label{p22}
\frac{\partial\pi_A}{\partial p_B}=\left(
\begin{array}{cc}
G'_{\alpha A}\tilde M'^{A\beta} & G'_{\alpha A}\tilde M'^{A j}  \\
G'_{i\alpha}& G'_{ij}
\end{array}\right) ~ = ~ \left(
\begin{array}{cc}
\tilde M'^{\alpha\beta} & \tilde M'^{\alpha j} \\
0 & \delta^{ij} 
\end{array}\right). 
\end{eqnarray}
Here we used that in adapted coordinates $G'_{\alpha A}=(\delta_{\alpha\beta}, {\bf 0})$ and $\tilde G'_{D \beta}=(\delta_{\alpha\beta}, {\bf 0})^T$. As $\tilde M'^{AB}$ is positive-definite matrix (see (\ref{int11.11})), we have $\det\tilde  M'^{\alpha\beta}>0$. Together with (\ref{p22}), this implies $\det(\partial\pi_A/\partial p_B)\ne 0$.

Representing $p_A$ through $q^A$ and $\pi_B$, we can rewrite the theory in terms of new variables. Our second-class constraints in the new coordinates are
\begin{eqnarray}\label{p23.1}
G_\alpha (q)=0, \qquad \pi_\alpha=0.
\end{eqnarray}
Using the canonical brackets $\{q^A, p_B\}=\delta^A{}_B$, we get the following Poisson brackets for  the variables $q^A$ and $\pi_i$
\begin{eqnarray}\label{p24}
\{ q^A, q^B\}=0, \qquad \{q^\alpha, \pi_i \}=G_{i\alpha}(q), \qquad \{q^j, \pi_i \}=G_{ij}(q)=\delta_{ij}, \qquad 
\{\pi_i, \pi_j \}=-c_{ij}{}^D(q)p_D(q^A, \pi_i, \pi_\alpha), 
\end{eqnarray}
where appeared the Lie brackets of basic vector fields ${\bf G}_i$
\begin{eqnarray}\label{p25}
c_{ij}{}^D=[{\bf G}_i, {\bf G}_j]^D\equiv G_{iE}\partial_E G_{jD}-G_{jE}\partial_E G_{iD}. 
\end{eqnarray}
As above, the special property of new variables is that  $q^A$ and $\pi_i$ have vanishing brackets with the original constraints $G_\alpha$
\begin{eqnarray}\label{p23}
\{ q^A, G_\alpha(q) \}=0, \qquad \{\pi_i, G_\alpha(q) \}=\{G_{i A}(q)p_A, G_\alpha(q) \}=-G_{i A}G_{\alpha A}=0, 
\end{eqnarray}
the latter equality is due to Eq. (\ref{p9}).  On this reason, when we pass to the Dirac bracket, the brackets (\ref{p24}) will not be modified, while the 
brackets of $\pi_\alpha=\Phi_\alpha$ with any phase-space function vanish.  The final Hamiltonian is obtained from (\ref{p3.3}) disregarding the last two terms and substituting $p_A(q^A, \pi_i, \pi_\alpha=0)$ into the remaining terms
\begin{eqnarray}\label{p24.1}
H'_0(q^A, \pi_j)=p_A(q^A, \pi_j, 0)v^A(q^B, p_C(q^A, \pi_j, 0))-L(q^B, p_C(q^A, \pi_j, 0)). 
\end{eqnarray}
The final brackets are (\ref{p24}), where we substitute $\pi_\alpha=0$ on r.h.s. of the last equation. 
Hamiltonian equations are obtained with use of the final brackets as follows: $\dot q^A=\{q^A, H'_0( q^B, \pi_j) \}'$, $\dot\pi_i=\{\pi_i, H'_0( q^B, \pi_j) \}'$.

\section{Integration of first-order equations with use of Hamiltonian vector field.}\label{CAP1.6}
To apply in practice the Liouvolle's theorem discussed in Sect. \ref{CAP1.2}, we need to find the integrals of motion, then solve the algebraic equations (\ref{int16}), then we need to calculate the integrals given in Eqs. (\ref{int30}) (for the rigid body they typically are the elliptic integrals), and finally, solve the algebraic equations (\ref{int30}). In this section we present another possibility to integrate first-order equations in terms of power series with respect to $t$. 

Consider the differential operator acting on space of functions $f(x)$ and defined by formal series
\begin{eqnarray}\label{hvf1}
e^{h\partial_x} = 1 + h\partial_x + \frac{1}{2}h\partial_x (h\partial_x ) + \ldots = \sum\limits_{n = 0}^\infty {\frac{1}{n!}}
(h\partial_x )^n.  
\end{eqnarray}
where $h = \mbox{const}$, and $\partial_x=\frac{\partial}{\partial x}$. This obeys the properties $e^{h\partial_x }\,x
= x+h$, $e^{h\partial_x }f(x) = f(e^{h\partial_x }x)$, as can be verified by expansion in power series of both sides
of these equalities. There is a generalization of the last equality for the case of a function $h(x)$. For the latter use we introduce the parameter $t$. Then
\begin{eqnarray}\label{hvf2}
e^{th(x)\partial_x }f(x) = f(e^{th(x)\partial_x}x), \qquad \mbox{in particular} \qquad e^{th(x)\partial_x }h(x) = h(e^{th(x)\partial_x}x).
\end{eqnarray}
To prove this\footnote{This proof was suggested by Andrey Pupasov-Maksimov.}, let us consider the following Cauchy problem for partial differential equation
\begin{eqnarray}\label{hvf3}
\partial_t\varphi(t, x)=h(x)\partial_x\varphi(t, x), \qquad  \varphi(0, x)=f(x), 
\end{eqnarray}
where $h(x)$ and $f(x)$ are given functions. It is known (see Sect. 60 in \cite{Pet_1973}), that this problem has unique solution $\varphi(t, x)$. The 
function $e^{th(x)\partial_x }f(x)$ obeys to this problem
\begin{eqnarray}\label{hvf4}
\partial_t[e^{th(x)\partial_x}f(x)]=\partial_t[1+th\partial_x+\frac{t^2}{2!}(h\partial_x)^2+\ldots]f(x)= \cr 
[h\partial_x+t(h\partial_x)^2+\frac{t^2}{2!}(h\partial_x)^3+\ldots]f(x)=h\partial_x[e^{th(x)\partial_x}f(x)]. 
\end{eqnarray}
Denoting $e^{th(x)\partial_x}x\equiv y(x)$, we verify that the function $f(e^{th(x)\partial_x}x)$ also obeys to this problem
\begin{eqnarray}\label{hvf5}
\partial_t[f(e^{th(x)\partial_x}x)]=\left.\partial_y f\right|_{y(x)}\partial_t[e^{th(x)\partial_x}x]=\left.\partial_y f\right|_{y(x)}\partial_t[1+th\partial_x+\frac{t^2}{2!}(h\partial_x)^2+\ldots]x= \cr 
\left.\partial_y f\right|_{y(x)}[h\partial_x+t(h\partial_x)^2+\frac{t^2}{2!}(h\partial_x)^2+\ldots]x=
\left.\partial_y f\right|_{y(x)}h\partial_x[1+t(h\partial_x)+\frac{t^2}{2!}(h\partial_x)^2+\ldots]x= \cr 
h\left.\partial_y f\right|_{y(x)}\partial_x y(x)=h\partial_x[f(y(x))]=
h\partial_x[f(e^{th(x)\partial_x}x)]. \qquad 
\end{eqnarray}
Since the solution is unique, the two functions must coincide, which proves the equality (\ref{hvf2}). 

As a consequence, the series $z(t , x) = e^{th(x)\partial_x}x$ turns
out to be a general solution to the equation
\begin{eqnarray}\label{hvf6}
\dot z = h(z),
\end{eqnarray}
with $x$ being the integration constant. 
Indeed, we have 
\begin{eqnarray}\label{hvf7}
\dot z=\frac{d}{dt}[e^{th(x)\partial_x}x] =\partial_t[1+th\partial_x+\frac{t^2}{2!}(h\partial_x)^2+\ldots]x=[h\partial_x+t(h\partial_x)^2+\frac{t^2}{2!}(h\partial_x)^3+\ldots]x= \cr 
h+t(h\partial_x)h+\frac{t^2}{2!}(h\partial_x)^2h+\ldots=e^{th(x)\partial_x}h(x)=h(e^{th(x)\partial_x}x)=h(z), \qquad \qquad 
\end{eqnarray}
where the penultimate equality is due to (\ref{hvf2}).

This observation immediately generalized for the case of several variables: the functions 
\begin{eqnarray}\label{hvf7.1}
z^i(t , z_0^j) = e^{t h^k(z_0^j)\frac{\partial}{\partial z_0^k}}z_0^i, 
\end{eqnarray}
provide a general solution to the system
\begin{eqnarray}\label{hvf8}
\dot z^i = h^i(z^j). 
\end{eqnarray}
Any Hamiltonian system $\dot x^i = \{x^i, H\}$, $\dot p_j = \{p_j, H\}$ has this form. So its general solution is
\begin{eqnarray}\label{hvf9}
x^i(t, x^j_0, p_{0 k}) = e^{t \{x_0^k ,H(x_0, p_0)\}\frac{\partial }{\partial x_0^k }+t \{p_{0 k} ,H(x_0, p_0)\}\frac{\partial }{\partial p_{0k}}}x_0^i, \cr  
p_i(t, x^j_0, p_{0 k}) = e^{t \{x_0^k ,H(x_0, p_0)\}\frac{\partial }{\partial x_0^k }+t \{p_{0 k} ,H(x_0, p_0)\}\frac{\partial }{\partial p_{0k}}}p_{0 i}.  
\end{eqnarray}
There is a generalization of these formulas to the case of time-dependent Hamiltonian, see \cite{deriglazov2010classical}.

\section{Application of intermediate formalism to a toy model.}\label{CAP1.6.1}

Here we illustrate the intermediate formalism on the example of a particle on a sphere, obtaining a non standard Hamiltonian description of these model in five-dimensional symplectic manifold.

Consider a point particle with coordinates  $x_i(t)$ in three-dimensional Euclidean space, forced to move freely on the sphere ${\bf x}^2=c^2$. It can be described by the Lagrangian action
\begin{eqnarray}\label{dm1}
S=\int dt ~ \frac{m}{2}\dot{\bf x}^2+\lambda({\bf x}^2-c^2).
\end{eqnarray}
In the phase space with canonically conjugated coordinates $({\bf x}$, ${\bf p})$, this action implies two second-class constraints ${\bf x}^2-c^2=0$ 
and $({\bf x}, {\bf p})=0$. The first is analogous to $G_\alpha=0$ of general formalism, while the second is analogous of $\Phi_\alpha=0$ and determines five-dimensional intermediate submanifold in the phase space. Then the analogous of $G_{\alpha A}$ is the vector $\frac12 {\bf grad}({\bf x}^2-c^2)=(x_1, x_2, x_3)$. Assuming that we work in the local coordinate chart with $x_3\ne 0$, fundamental solutions to the equation $({\bf x}, {\bf z})=0$ 
are $(1, 0, -\frac{x_1}{x_3})$ and $(0, 1, -\frac{x_2}{x_3})$. The change of variables (\ref{p10}) reads
\begin{eqnarray}\label{dm2}
\left(
\begin{array}{c}
\pi_1 \\
\pi_2 \\
\pi_3
\end{array}\right)=\left(
\begin{array}{ccc}
1, & 0 & -\frac{x_1}{x_3} \\
0 & 1 & -\frac{x_2}{x_3} \\
x_1 & x_2 & x_3
\end{array}\right)\left(
\begin{array}{c}
p_1 \\
p_2 \\
p_3
\end{array}\right). 
\end{eqnarray}
We get
\begin{eqnarray}\label{dm3}
\pi_1=p_1-\frac{x_1}{x_3} p_3, \qquad \pi_2=p_2-\frac{x_2}{x_3} p_3, \qquad \pi_3=({\bf x}, {\bf p}). 
\end{eqnarray}
Hence in the new coordinates ${\bf x}$ and ${\boldsymbol\pi}$, the intermediate submanifold is just the hyperplane $\pi_3=0$. The inverse  transformation to (\ref {dm3}) is 
\begin{eqnarray}\label{dm4}
p_1=\pi_1-\frac{x_1}{{\bf x}^2}\left[x_1\pi_1+x_2\pi_2-\pi_3\right], \qquad p_2=\pi_2-\frac{x_2}{{\bf x}^2}\left[x_1\pi_1+x_2\pi_2-\pi_3\right], \qquad  p_3=-\frac{x_3}{{\bf x}^2}\left[x_1\pi_1+x_2\pi_2-\pi_3\right].
\end{eqnarray}
The next step is to rewrite the canonical Poisson brackets $\{x_i, p_j\}=\delta_{ij}$ and Hamiltonian $H=\frac{1}{2m}{\bf p}^2$ in terms of new coordinates, and then substitute $\pi_3=0$ in all resulting expressions. Using the expressions  (\ref{dm2}) and the canonical brackets, we obtain the following non vanishing brackets for the coordinates $(x_1, x_2, x_3, \pi_1, \pi_2)$ of intermediate submanifold
\begin{eqnarray}\label{dm5}
\{x_1, \pi_1\}=1, \qquad \{x_2, \pi_2\}=1, \qquad \{x_3, \pi_1\}=-\frac{x_1}{x_3}, \qquad \{x_3, \pi_2\}=-\frac{x_2}{x_3}. 
\end{eqnarray}
They do not involve $\pi_3$, so they already give the Poisson structure of intermediate manifold. Using Eqs. (\ref{dm4}) in the canonical Hamiltonian, and then setting  $\pi_3=0$, we obtain the Hamiltonian reduced to the intermediate submanifold 
\begin{eqnarray}\label{dm6}
H=\frac{1}{2m}\left\{\pi_1^2+\pi_2^2-\frac{(x_1\pi_1+x_2\pi_2)}{{\bf x}^2}\right\}. 
\end{eqnarray}
Eqs. (\ref{dm5}) and (\ref{dm6}) represent a Hamiltonian system on a five-dimensional symplectic manifold foliated by the 
leaves ${\bf x}^2=c^2$, $c\in{\mathbb R}$.  
The quantity  ${\bf x}^2$ is a Casimir function of the Poisson structure (\ref{dm5}). So any trajectory that passes through a point of the  symplectic 
leaf ${\bf x}^2=c^2$ with given $c$, entirely lies in this leaf.

\section{Rotating asymmetric body in the intermediate formalism.}\label{CAP1.7}

Here we apply the intermediate formalism to a spinning body.  We show that Euler-Poisson equations turn out to be a Hamiltonian system on the intermediate submanifold, and deduce the Poisson geometry (\ref{11.20}) that lie behind these equations.

Motions of a spinning body can be described \cite{AAD23_1,AAD23} starting from the Lagrangian action of the form (\ref{int02})
\begin{eqnarray}\label{irb1}
S=\int dt ~ ~ \frac12 g_{ij}\dot R_{ki}\dot R_{kj} -\frac12 \lambda_{ij}\left[R_{ki}R_{kj}-\delta_{ij}\right],
\end{eqnarray}
where $R^TR-{\bf 1}$ play the role of $G_\alpha$ of the general formalism. 
The action is written in Laboratory system with the origin chosen at center of mass of the body. $R_{ij}(t)$ is $3\times 3$ matrix. Its nine elements are the dynamical degrees of freedom which, at the end, describe rotational motions of the body.  The numeric symmetric matrix $g_{ij}$ encodes the distribution of mass of the body at initial instant
\begin{eqnarray}\label{irb2}
g_{ij}\equiv\sum_{N=1}^{n}m_Nx_N^i(0)x_N^j(0),
\end{eqnarray}
where $m_N$ are masses of the body's particles with position vectors ${\bf x}_N(t)$.  The mass matrix and inertia tensor are related as 
follows: $I_{ij}=g_{kk}\delta_{ij}-g_{ij}$. Choosing Laboratory axes at $t=0$ in the directions of axes of inertia, the two tensors acquire a diagonal form, $g_{ij}=g_i\delta_{ij}$, $I_{ij}=I_i\delta_{ij}$. For a non planar body, $g_i$ are positive 
numbers \cite{AAD23}, so the Hessian matrix of the theory (\ref{irb1}) evidently is positive-defined. Therefore we can apply the intermediate formalism developed  in Sects. \ref{CAP1.4} and \ref{CAP1.5}. 

Introducing the conjugate momenta for all dynamical variables: $p_{ij}=\partial L/\partial\dot R_{ij}$ and $p_{\lambda ij}=\partial L/\partial\dot\lambda_{ij}$, we obtain the expression for $p_{ij}$ in terms of velocities
\begin{eqnarray}\label{3.2}
p_{ij}=\dot R_{ik}g_{kj}, \quad \mbox{then} ~ \dot R_{ij}=p_{ik}g^{-1}_{kj},
\end{eqnarray}
and the primary constraints $p_{\lambda ij}=0$.
To construct the final Hamiltonian of the intermediate formalism, we will need only canonical part $H_0=p_{ij}\dot R_{ij}(p)-L(\dot R_{ij}(p))$ of the complete Hamiltonian (\ref{p3.3}). For the present case, its explicit form is 
\begin{eqnarray}\label{3.4}
H=\frac12 g^{-1}_{ij}p_{ki}p_{kj}.
\end{eqnarray}
The non vanishing Poisson brackets of canonical variables are (there is no summation over $i$ and $j$): $\{ R_{ij}, p_{ij}\}=1$, $\{\lambda_{ij}, p_{\lambda ij}\}=1$. 
Next, the explicit form of tertiary constraints (\ref{p6}) in our case is 
\begin{eqnarray}\label{3.6}
\{R_{ki}R_{kj}, H_0\}=\left[R^Tpg^{-1}+(R^Tpg^{-1})^T\right]_{ij}\equiv\left[R^TRR^{-1}pg^{-1}+(R^TRR^{-1}pg^{-1})^T\right]_{ij}=0. 
\end{eqnarray}
The surface determined by equations $R^TR={\bf 1}$ and (\ref{3.6}) is equally determined by $6+6$  equations
\begin{eqnarray}
R^TR={\bf 1}, \qquad \qquad \qquad \label{3.4.1} \\
\Phi_{ij}\equiv\left[R^{-1}pg^{-1}+(R^{-1}pg^{-1})^T\right]_{ij}=0. \label{3.4.2}
\end{eqnarray}
We take these $\Phi_{ij}$ as analogues of the constraints (\ref{p6})  of the general formalism. 

According to the intermediate formalism, we now need to find non-canonical momenta with two properties. First, $9-6=3$ of them should have vanishing Poisson brackets with the orthogonality constraint, see Eq. (\ref{p11}).   Second, the constraints (\ref{3.4.2}) can be used to represent other momenta through these three, see Eq. (\ref{p19.0}). To achieve this, consider the phase-space functions
\begin{eqnarray}\label{3.6.1}
{\mathbb P}_{ij}\equiv 2(R^{-1}p)_{ij}, \qquad \mbox{then} \quad p_{ij}=\frac12 (R{\mathbb P})_{ij}.  
\end{eqnarray}
They are constructed from $p_{ij}$ with use of invertible matrix, so the transition $(R_{ij}, p_{ij})\rightarrow(R_{ij}, {\mathbb P}_{ij})$ is a change of variables on the phase space. We emphasize that $R_{ij}$ in the action (\ref{irb1}) is an arbitrary (not orthogonal!) matrix. 

We  decompose ${\mathbb P}_{ij}$ on symmetric and antisymmetric parts, ${\mathbb P}_{ij}=S_{ij}-\hat M_{ij}$, where $S=R^{-1}p+(R^{-1}p)^T$ 
and $\hat M=R^{-1}p-(R^{-1}p)^T$, and then replace the antisymmetric matrix $\hat M$ on an equivalent vector\footnote{The phase-space functions $M_k$, being rewritten back in terms of $R_{ij}$ and $\dot R_{ij}$ are just the components of angular momentum in the body-fixed frame \cite{AAD23}: $M_k=R^{-1}_{ki}m_i$, $m_i=\sum_{N}m_N[{\bf x}_N(t), \dot{\bf x}_N(t)]_i$.}: $\hat M_{ij}=\epsilon_{ijk}M_k$, 
$M_k\equiv\frac12 \epsilon_{kij}\hat M_{ij}=-\epsilon_{kij}(R^{-1}p)_{ij}$. So the final form of the decomposition is 
\begin{eqnarray}\label{3.7}
{\mathbb P}_{ij}=S_{ij}-\epsilon_{ijk}M_k, \qquad \mbox{where} \quad S_{ij}=\left[R^{-1}p+(R^{-1}p)^T\right]_{ij}, \quad 
M_k=-\epsilon_{kij}(R^{-1}p)_{ij}. 
\end{eqnarray}
In accordance to this, we consider the following change of variables: 
\begin{eqnarray}\label{3.7.1}
(R_{ij}, ~ p_{ij}) \rightarrow (R_{ij},  ~ S_{ij},  i\le j, ~  M_k). 
\end{eqnarray}
The coordinates $M_k$ have the desired properties: their brackets with orthogonality constraint vanish: $\{ M_k, R_{pi}R_{pj}-\delta_{ij}\}=0$, and the variables $S_{ij}$ can be presented through $M_k$ resolving (\ref{3.4.2}) as follows (there is no summation on $i$ and $j$ in this expression)
\begin{eqnarray}\label{3.7.2}
S_{ij}=\frac{g_i-g_j}{g_i+g_j}\epsilon_{ijk}M_k=\frac{I_j-I_i}{I_k}\epsilon_{ijk}M_k. 
\end{eqnarray}
Therefore, the change of variables (\ref{3.7.1}) is analogous of the change (\ref{p10}) of the general formalism.
To obtain the last equality, we used the following relations among elements of diagonal mass matrix and inertia tensor \cite{AAD23}:
\begin{eqnarray}\label{1.35.1}
2g_1=I_2+I_3-I_1, \quad 2g_2=I_1+I_3-I_2, \quad 2g_3=I_1+I_2-I_3, \cr
I_1=g_2+g_3, \quad I_2=g_1+g_3, \quad I_3=g_1+g_2,  \qquad \qquad \cr
g_i-g_j=I_j-I_i. \qquad \qquad  \qquad \qquad \qquad
\end{eqnarray}

Computing the canonical Poisson brackets of the new variables $R_{ij}, M_k$ and $S_{ij}$ we get
\begin{eqnarray}\label{11.20}
\{ R_{ij}, R_{ab}\}=0, \qquad \{M_i, M_j\}=-\epsilon_{ijk}((R^TR)^{-1}{\bf M})_k, \qquad 
\{M_i, R_{jk}\}=-\epsilon_{ikm}R^{-1 T}_{jm};  
\end{eqnarray}
\begin{eqnarray}\label{11.21}
\{ R_{ij}, S_{ab}\}=R^{-1 T}_{ia}\delta_{jb}+R^{-1 T}_{ib}\delta_{ja}, \qquad 
\{ M_k, S_{ab}\}=-2M_k(R^TR)^{-1}_{ab}+\delta_{ka}((R^TR)^{-1}{\bf M})_b+\delta_{kb}((R^TR)^{-1}{\bf M})_a, \cr 
\{ S_{ij}, S_{ab}\}=-(R^TR)^{-1}_{ia}\epsilon_{jbn} M_n-(R^TR)^{-1}_{jb}\epsilon_{jan} M_n+(a\leftrightarrow b).  \qquad \qquad \qquad \qquad 
\end{eqnarray}

According to Sect. \ref{CAP1.5}, to reduce our theory on the submanifold $\Phi_{ij}=0$, it is sufficient to rewrite it in the variables $R_{ij}$, $M_k$, $S_{ij}$ and then, using Eq. (\ref{3.7.2}), to exclude from all resulting expressions the variables $S_{ij}$. The brackets (\ref{11.20}) do not involve $S_{ij}$. So they already give a Poisson structure of intermediate submanifold $\Phi_{ij}=0$. Using Eqs. (\ref{3.6.1}) and (\ref{3.7}) in the canonical Hamiltonian (\ref{3.4}), the latter can be written as follows:
\begin{eqnarray}\label{11.26.1}
H_0=\frac18 g^{-1}_{ij}(S_{ai}-\epsilon_{aik}M_k)(S_{aj}-\epsilon_{ajp}M_p)+\frac18 g^{-1}_{ij}(R^TR-{\bf 1})_{ab}(S_{ai}-
\epsilon_{aik}M_k)(S_{bj}-\epsilon_{bjp}M_p).
\end{eqnarray}
Second term is proportional to the orthogonality constraint. Therefore it does not contribute into Hamiltonian equations for the variables $R_{ij}$ and $M_k$, and hence it can be omitted. The remaining term can be written as follows:
\begin{eqnarray}\label{11.26.2}
H_0=\frac18 \sum_j \frac{1}{g_j}(S_{ij}-\epsilon_{ijk}M_k)^2. 
\end{eqnarray}
Using the relations (\ref{3.7.2}), for any chosen $i\ne j$ we get $\frac{1}{g_j}(S_{ij}-\epsilon_{ijk}M_k)^2=\frac{4g_j}{I^2_k}M^2_k$. Using this in (\ref{11.26.2}), we obtain final form of our Hamiltonian on the intermediate submanifold 
\begin{eqnarray}\label{11.26}
H_0=\frac12 I^{-1}_{ij}M_i M_j.    
\end{eqnarray}
Note that the final expression, being composed of tensors and vectors, is invariant under rotations. Hence the Hamiltonian will be of this form in any Laboratory system. If the Laboratory frame was not adapted with the axes of inertia at initial instant, the inertia tensor in this expression will be a numerical symmetric matrix with non-vanishing off-diagonal elements.

Using this $H_0$  with the brackets (\ref{11.20}), 
the Hamiltonian equations $\dot z=\{z, H_0\}$ read as follows: 
%\begin{eqnarray}\label{p26} 
$\dot R_{ij}=-\epsilon_{jkm}(I^{-1}M)_k R_{im}$, 
%\end{eqnarray}
%\begin{eqnarray}\label{p27}
$\dot{\bf M}=[{\bf M}, I^{-1}{\bf M}]$.   
%\end{eqnarray}
Introducing the phase-space quantity $\Omega_i=I^{-1}_{ij}M_j$, they acquire the standard form of Euler-Poisson 
equations: 
\begin{eqnarray}\label{p26}
\dot R_{ij}=-\epsilon_{jkm}\Omega_k R_{im},  \qquad  I\dot{\boldsymbol\Omega}=[I{\boldsymbol\Omega}, {\boldsymbol\Omega}]. 
\end{eqnarray}
By this, we completed Hamiltonian reduction on intermediate submanifold (\ref{3.4.2}), showing that Euler-Poisson equations are the Hamiltonian system on this submanifold, with the Poisson structure given by the brackets (\ref{11.20}).

{\bf The Chetaev bracket is the Dirac bracket.}\label{Chet} Using the orthogonality constraint on r.h.s. of the brackets (\ref{11.20}), we obtain more simple expressions
\begin{eqnarray}\label{11.20.2}
\{ R_{ij}, R_{ab}\}=0, \qquad \{M_i, M_j\}=-\epsilon_{ijk}M_k, \qquad 
\{M_i, R_{jk}\}=-\epsilon_{ikm}R_{jm}.  
\end{eqnarray}
By direct computations, it can be verified that they still satisfy the Jacobi identity and lead to the same equations (\ref{p26}).  They  were suggested by Chetaev \cite{Chet_1941} as the possible Poisson structure corresponding to the Euler-Poisson equations. 

{\bf General solution to the Euler-Poisson equations and the motions of a rigid body.} Not all solutions to the equations (\ref{p26}) describe the motions of a spinning body. By construction \cite{AAD23,AAD23_1}, they should be solved with the universal initial conditions
\begin{eqnarray}\label{11.20.1}
R_{ij}(0)=\delta_{ij}, \qquad \Omega_i(0)=\Omega_{0 i}, 
\end{eqnarray}
where $\Omega_{0 i}$ is initial angular velocity measured in the body-fixed frame. That is only those trajectories that at some instant of time pass through the unit of $SO(3)$\,-group can describe possible motions of the body\footnote{Misunderstanding of this point leads to a lot of confusion, 
see \cite{AAD23_2}.}.  Let us denote  the r.h.s. of equations (\ref{p26}) as follows: $H_{ij}(R, {\boldsymbol\Omega})$ and $H_k({\boldsymbol\Omega})$. Then, according to Eqs. (\ref{hvf7.1}), we can write for their general solution 
\begin{eqnarray}\label{dt61}
R_{ij}(t, R_{0 kp}, \Omega_{0 k})=e^{tH_{kp}(R_0, {\boldsymbol\Omega}_0)\frac{\partial}{\partial R_{0 kp}}+tH_k({\boldsymbol\Omega}_0)\frac{\partial}{\partial\Omega_{0 k}}}R_{0ij}. 
\end{eqnarray}
After applying the differential operator in the exponential, $R_{0 kp}$ should be replaced on $\delta_{kp}$ in each term of the obtained power series.
The resulting function $R_{ij}(t, \Omega_{0 k})$ will represent the motion of a spinning body, that at $t=0$ has its inertia axes parallel with the Laboratory axes, and the initial angular velocity equal to $\Omega_{0 i}$.

\section{Conclusion.}

The most economical Hamiltonian formulation of the theory (\ref{int02}), in which we are interested in knowing the dynamics of all variables $q^A$, is achieved on the intermediate submanifold of phase space determined by the constraints (\ref{p6}) (or, equivalently, by (\ref{p19.0})).  We have described and discussed two ways of Hamiltonian reduction to this submanifold. The final result of the reduction using the Dirac bracket is written out in equations (\ref{p8.13})-(\ref{p8.15}). The intermediate formalism gives the equations (\ref{p20})-(\ref{p14.3}). As we have shown in the last section, namely the intermediate formalism directly leads to the Euler-Poisson equations of a spinning body. 

To further compare the two reductions, let us denote coordinates $(q^A, p_B)$ of original phase space by $z^i$, while coordinates $(q^A, p_j)$ of intermediate submanifold by $z^a$. Let the Poisson tensor of original space is $\omega^{ij}$, the Dirac tensor is $\omega^{ij}_D$ and the Poisson tensor induced on intermediate submanifold  is $\bar\omega^{ab}$. Generally, in the process of reduction with use of Dirac bracket, $\omega^{ij} \rightarrow \omega^{ij}_D \rightarrow \bar\omega^{ab}$, we have $\omega^{ab}\ne\omega^{ab}_D\ne\bar\omega^{ab}$. An alternative possibility, developed in this work, can be resumed as follows: in the theory (\ref{int02}) with a positive-definite Lagrangian $L$, there are phase-space coordinates $z'^i=(q^A, \pi_B)$ such that in the process of reduction $\omega'^{ij} \rightarrow \omega'^{ij}_D \rightarrow \bar\omega'^{ab}$ we have $\omega'^{ab}=\omega'^{ab}_D\ne\bar\omega'^{ab}$. 
In view of this, the reduction consists in exclusion of redundant momenta (see Eq. (\ref{p19.0})) from the block $\omega'^{ab}$ of original tensor $\omega'^{ij}$.

%{\bf Acknowledgments.} 
\begin{acknowledgments}
The work has been supported by the Brazilian foundation CNPq (Conselho Nacional de
Desenvolvimento Cient\'ifico e Tecnol\'ogico - Brasil).
\end{acknowledgments}

\section{Appendix.}\label{App_1}
Here we enumerate some properties of a positive-definite matrix. 

Symmetric real-valued $n\times n$\,-matrix  with the elements $M_{ij}$ is called positive-definite ($M\succ 0$), if for any  non zero column ${\bf Y}$ we have ${\bf Y}^TM{\bf Y}>0$. The following affirmations turn out to be equivalent \cite{Shi_1977}: \par

\noindent {\bf 1A.} $M\succ 0$. \par

\noindent {\bf 1B.} There exists $n\times n$ positive-definite matrix $B$ such that $M=B^2\equiv B^T B$. \par

\noindent {\bf 1C.} All principal minors of $M$ are positive numbers. In particular, $\det M>0$. \par

\noindent {\bf 1D.} $M$ is the Gram matrix of some set of $p$\,-dimensional linearly independent vectors, say ${\bf Z}_i$. That 
is $M_{ij}=({\bf Z}_i, {\bf Z}_j)$. If $Z_{Ai}$, $A=1, 2, \ldots p$ is the matrix formed by the columns ${\bf Z}_i$, we can write $M_{ij}=(Z^T)_{iA}Z_{Aj}$. \par

\noindent {\bf 1E.} All eigenvalues of $M$ are positive numbers. 

Besides, there are the following properties: \par

\noindent {\bf 2A.} Diagonal elements of positive-definite matrix are positive numbers: $M_{ii}>0$ for any $i$. Then $\mbox{trace} ~ M>0$. \par

\noindent {\bf 2B.} Positive-definite matrix is invertible, and its inverse is a positive-definite matrix. 

{\bf Affirmation.} Let $\mbox{rank} ~ Q_{Ai}=k$, where $A=1, 2, \ldots p$, $i=1, 2, \ldots k$, $k<p$, and $M_{AB}$ is positive-definite. Then the matrix
\begin{eqnarray}\label{app01}
N_{ij}=(Q^T)_{iA}M_{AB}Q_{Bj}, 
\end{eqnarray}
is non-degenerate, $\det N\ne 0$.

{\it Proof.} Using {\bf 1B}, we write $M=B^TB$, then  
\begin{eqnarray}\label{app02}
N_{ij}=(BQ)^T_{iA}(BQ)_{Aj}, 
\end{eqnarray}
where, according to {\bf 2B}, the matrix $B$ is nondegenerate. Since the columns of $Q_{Aj}$ are linearly independent, the matrix $(BQ)_{Aj}$ also is composed of linearly independent columns. Then (\ref{app02}) means that $M_{ij}$ is the Gram matrix. According to {\bf 1D} it is positive-definite. In particular, $\det N>0$.

\end{document}